# $\mathcal{P} \cdot \mathcal{T} \cdot \mathcal{D}$ Symmetry Protected Scattering Anomaly in Optics

Mário G. Silveirinha[*]

[(1)]University of Lisbon–Instituto Superior Técnico and Instituto de Telecomunicações, Avenida Rovisco Pais, 1, 1049-001 Lisboa, Portugal

## Abstract

In time-reversal invariant electronic systems the scattering matrix is anti-symmetric. This property enables an effect, designated here as "scattering anomaly", such that the electron transport does not suffer from back reflections, independent of the specific geometry of the propagation path or the presence of time-reversal invariant defects. In contrast, for a generic time-reversal invariant photonic system the scattering matrix is symmetric and there is no similar anomaly. Here, it is theoretically proven that despite these fundamental differences there is a wide class of photonic platforms – in some cases formed only by time-reversal invariant media – in which a scattering anomaly can occur. It is shown that an optical system invariant under the action of the composition of the parity, time-reversal, and duality operators ($\mathcal{P} \cdot \mathcal{T} \cdot \mathcal{D}$) is characterized by an anti-symmetric scattering matrix. Specific examples of photonic platforms wherein the scattering anomaly occurs are given, and it is demonstrated with full wave numerical simulations that the proposed systems enable bidirectional waveguiding immune to arbitrary deformations of the propagation path. Importantly, our theory unveils a new class of fully three dimensional structures wherein the transport of light is fully protected against reflections and uncovers unsuspected links between the electrodynamics of reciprocal and nonreciprocal materials.

---

[*] To whom correspondence should be addressed: E-mail: *mario.silveirinha@co.it.pt*

# I. Introduction

In recent years there has been a great interest in different paradigms for the transport of light immune to the unwanted effects of reflections created by imperfections, disorder, obstacles or other types of deformations of the propagation path [1, 2]. In particular, a lot of attention has been devoted to the realization of novel topological phases that may allow for the flow of optical energy with no backscattering. An important class of solutions is based on (nonreciprocal) media with a broken time reversal-symmetry [3-13], usually created by a biasing static magnetic field. It has been shown that such platforms are inherently topological and characterized by nontrivial Chern indices [3, 4, 12]. In parallel, different research groups explored alternative solutions that can offer some form of topological protection using only time-reversal invariant materials. Different concepts were developed based on these efforts: Floquet topological insulators [14-15], systems with a synthetic pseudo-magnetic field [16-17], photonic topological insulators [18-24], photonic crystals and waveguides with specific spatial symmetries [25, 26], and other related ideas [27, 28].

In contrast with most of previous works here, rather than focusing on the topological properties of a system, we investigate the conditions under which a photonic platform can have propagation channels insensitive to backscattering. Our study is inspired by the fact that in electronics a system that is invariant under the time-reversal operation is characterized by an anti-symmetric scattering matrix. In some circumstances, this property creates the conditions for a "scattering anomaly" such that an electron wave can propagate through rather complex meandering paths totally insensitive to the effects of disorder, defects or path deformation [29]. In electronics, this property is intimately linked to a $\mathbb{Z}_2$ topological invariant and to the theory of electronic topological insulators [29, 30, 31].

Different from the electronic case, in optics the scattering matrix is known to be symmetric for reciprocal systems and unfortunately there is no "scattering anomaly" associated with such a type of symmetry. Despite this difficulty, we prove here that there is a wide class of photonic platforms wherein the scattering anomaly can occur. It is shown that systems invariant under the action of a $\mathcal{P}\cdot\mathcal{T}\cdot\mathcal{D}$ operator – the composition of the parity, time-reversal, and duality transformations – are characterized by an anti-symmetric scattering matrix, similar to the electronic case. Specific examples of such systems are studied in detail and it is demonstrated with full wave simulations that the $\mathcal{P}\cdot\mathcal{T}\cdot\mathcal{D}$ symmetry protection may enable bidirectional waveguiding totally insensitive to back reflections. Furthermore, it is highlighted that the $\mathcal{P}\cdot\mathcal{T}\cdot\mathcal{D}$-invariance can be compatible with the time-reversal invariance, and hence our designs can rely on reciprocal materials.

Recently *He et al* constructed a fermionic type pseudo-time reversal operator to justify the absence of scattering in photonic topological insulators [24], in a spirit similar to what is done here. Yet, the operator of Ref. [24] is generally rather different from ours because it acts only on the fields and not on the spatial coordinates. Indeed, it is possible to construct different fermionic-type time reversal operators, and here we explore another solution using the parity operator as one of the elements of the transformation. Our theory generalizes and connects several previous works [18, 24, 26]. In particular, it includes as a particular case the photonic topological insulators discovered by A. B. Khanikaev, *et al* [18-24] using different ideas. Indeed, similar to Ref. [18] we find that reciprocal media with an Ω-type coupling may enable scatter-free waveguiding. The studies of Refs. [18-24] are focused on topological photonic crystals, different from ours that is based on generic electromagnetic continua and wherein the materials are not required to be topological. Indeed, even though often there are important connections between the "scattering anomaly" and topological concepts, the $\mathcal{P}\cdot\mathcal{T}\cdot\mathcal{D}$-invariance is inherently a *single-frequency* condition, and hence it is independent of the detailed material response away from the spectral region of interest. Thus, the requirements of our theory are very much relaxed as compared to those of topological theories, which usually enforce the system to be periodic and the materials response to have complicated symmetries in a broadband region. Furthermore, the theoretical concepts introduced in this article unveil totally different solutions for the transport of light not previously explored in the literature, and in particular uncover a way to realize *three-dimensional* (3D) optical edge waveguides fully protected against backscattering due to deformations of the propagation path. Remarkably, our general framework also includes as a particular case the paradigm discovered in Ref. [26] to have symmetry-protected light transport in a bulk waveguide – formed by standard isotropic dielectrics – with special boundary conditions linked by duality and parity. Thus, our analysis reveals an unexpected link between the bulk waveguides of Ref. [26] and the photonic topological insulator edge waveguides, showing that both correspond to particular cases of $\mathcal{P}\cdot\mathcal{T}\cdot\mathcal{D}$ invariant systems. Our theory merges and expands these two paradigms and in this manner opens new ways for waveguiding immune to backscattering.

## II. Theory

For notational convenience, we write the Maxwell's equations in the frequency domain as:

$$\hat{N} \cdot \mathbf{f} = \omega \mathbf{g}, \quad \text{with} \quad \hat{N} = \begin{pmatrix} \mathbf{0} & i\nabla \times \mathbf{1}_{3\times 3} \\ -i\nabla \times \mathbf{1}_{3\times 3} & \mathbf{0} \end{pmatrix}. \quad (1)$$

Here, $\mathbf{1}_{3\times 3}$ is the 3×3 identity matrix, $\omega$ is the oscillation frequency, $\mathbf{f} = (\mathbf{E} \ \mathbf{H})^T$ and $\mathbf{g} = (\mathbf{D} \ \mathbf{B})^T$ are six-component vector fields written in terms of the standard electromagnetic field vectors, and $T$ denotes the transpose operator. It is assumed that the $\mathbf{f}$ and $\mathbf{g}$ fields are linked as $\mathbf{g} = \mathbf{M}(\mathbf{r}) \cdot \mathbf{f}$ where $\mathbf{M}$ is a space-dependent material matrix of the generic bianisotropic form [32]:

$$\mathbf{M}(\omega) = \begin{pmatrix} \varepsilon_0 \bar{\varepsilon} & \frac{1}{c} \bar{\xi} \\ \frac{1}{c} \bar{\zeta} & \mu_0 \bar{\mu} \end{pmatrix}. \quad (2)$$

The tensors $\bar{\varepsilon}(\omega), \bar{\mu}(\omega), \bar{\xi}(\omega), \bar{\zeta}(\omega)$ are dimensionless and determine the permittivity, permeability and the magneto-electric coupling tensors, respectively.

In this compact formalism, the time-reversal operator $\mathcal{T}$ is defined by $\mathcal{T} = \mathcal{K} \boldsymbol{\sigma}_z$ where $\mathcal{K}$ denotes the complex conjugation operator and $\boldsymbol{\sigma}_z = \begin{pmatrix} \mathbf{1}_{3\times 3} & 0 \\ 0 & -\mathbf{1}_{3\times 3} \end{pmatrix}$. The time-reversal operation transforms the (frequency domain) electromagnetic fields as $\mathbf{f} \to \mathcal{T} \cdot \mathbf{f}$ and $\mathbf{g} \to \mathcal{T} \cdot \mathbf{g}$. The operator $\mathcal{T}$ is anti-linear, i.e., it is the composition of a linear operator and $\mathcal{K}$. Importantly, the action of $\mathcal{T}$ flips the Poynting vector.

As is well known, in optics the time-reversal operator satisfies $\mathcal{T}^2 = \mathbf{1}$ [1]. This contrasts sharply with the electronic case wherein $\mathcal{T}^2 = -\mathbf{1}$, a property that makes possible to have a "scattering anomaly". The anomaly occurs when the number of propagating states along a fixed direction, let us say the +*x*-direction, is odd [29]. What is extraordinary is that under these circumstances it is possible to have bidirectional waveguiding absolutely insensitive to any form of perturbation that preserves the time-reversal invariance of the system [29, 30, 31].

## *A.     The scattering matrix*

It would be highly interesting to have bidirectional waveguiding in optics insensitive to backscattering. This requires finding some anti-linear operator $\tilde{\mathcal{T}}$ with $\tilde{\mathcal{T}}^2 = -\mathbf{1}$ and that transforms the electromagnetic fields in such a manner that its action flips the Poynting vector in the relevant propagation directions, similar to the time-reversal operation. To demonstrate that a photonic platform invariant under the action of such hypothetical operator may enable scatter-free wave propagation we consider the scenario of Fig. 1, which represents two generic waveguides connected at some arbitrary junction corresponding to a perturbation of the system, e.g., a sinuous winding path that joins the two waveguides. Each waveguide supports a finite number of propagating modes at the frequency of interest. As is well known, sufficiently far from the junction – in particular the vicinity of the waveguide "ports" – it is possible to expand the electromagnetic field in terms of the propagating modes, because other (evanescent-type) modes play no role.

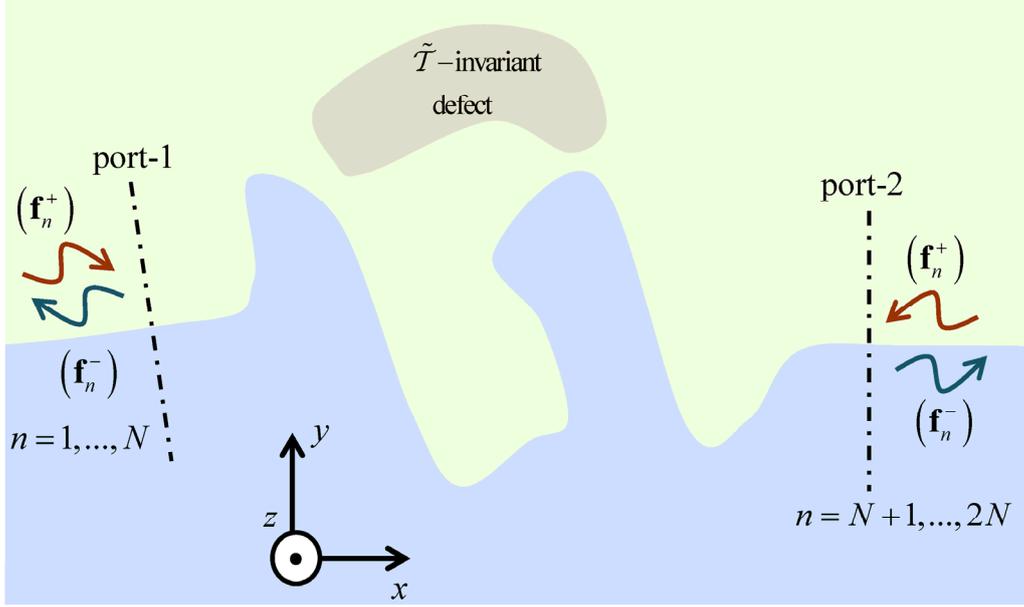

Fig. 1. (Color online) **Scattering in a $\tilde{\mathcal{T}}$-invariant photonic platform.** Two propagation channels (the edge waveguides) are connected by a meandering path with arbitrary deformations and with $\tilde{\mathcal{T}}$-invariant defects. Each channel supports a finite number of modes $N$ propagating towards the waveguide ports ($\mathbf{f}_n^+$). In case of a scattering anomaly, i.e., when $N$ is odd, and for lossless materials there is at least a propagation channel that allows an incoming wave to go through the sinuous path with no backscattering.

The "incident" wave that propagates towards the junction can be written as:

$$\mathbf{f}^+ = \begin{cases} \sum_{n=1}^{N} a_n^+ \mathbf{f}_n^+, & \text{waveguide 1} \\ \sum_{n=N+1}^{2N} a_n^+ \mathbf{f}_n^+, & \text{waveguide 2} \end{cases} \quad (3)$$

where $\mathbf{f}_n^+ = \mathbf{f}_n^+(\mathbf{r}; \omega)$ is the electromagnetic field associated with a given mode, $a_n^+$ is the corresponding complex-valued incident amplitude calculated at the relevant waveguide port, and $N$ is the number of propagating modes travelling towards the junction in each waveguide. The $\mathbf{f}_n^+$ modes are supposed to be normalized so that they transport the same power. It is also supposed that the two waveguides support the same number of propagating modes. Note that the indices $n=1,…N$ are associated with the modes of waveguide 1, whereas the indices $n=N+1,…2N$ are associated with the modes of waveguide 2. The incident modes originate scattered waves that propagate away from the junction. If the waveguides are invariant under the action of $\tilde{\mathcal{T}}$ the scattered waves can be expanded as follows:

$$\mathbf{f}^- = \begin{cases} \sum_{n=1}^{N} a_n^- \tilde{\mathcal{T}} \cdot \mathbf{f}_n^+, & \text{waveguide 1} \\ \sum_{n=N+1}^{2N} a_n^- \tilde{\mathcal{T}} \cdot \mathbf{f}_n^+, & \text{waveguide 2} \end{cases}. \qquad (4)$$

Indeed, because by hypothesis $\tilde{\mathcal{T}}$ flips the direction of the energy flow (Poynting vector) it is possible to pick $\mathbf{f}_n^- = \tilde{\mathcal{T}} \cdot \mathbf{f}_n^+$ ($n=1,…2N$) as the basis for the expansion of the scattered waves. Because of the linearity of the problem the scattered wave complex amplitudes can be linked to the incident wave amplitudes through a scattering matrix: $\left[ a_n^- \right] = \mathbf{S} \cdot \left[ a_n^+ \right]$. The scattering matrix can be written as $\mathbf{S} = \begin{pmatrix} \mathbf{S}_{11} & \mathbf{S}_{12} \\ \mathbf{S}_{21} & \mathbf{S}_{22} \end{pmatrix}$ where $\mathbf{S}_{ij}$ are $N \times N$ matrices. Crucially, if the entire structure (waveguides and junction) is invariant under the action of $\tilde{\mathcal{T}}$ then a solution of Maxwell's equations is transformed by $\tilde{\mathcal{T}}$ into another solution. Since $\tilde{\mathcal{T}}$ flips the Poynting vector, it transforms the incident wave $\mathbf{f}^+$ into a wave that propagates away from the junction ($\tilde{\mathcal{T}} \cdot \mathbf{f}^+$) and the scattered wave $\mathbf{f}^-$ into a wave that impinges on the junction. Because $\tilde{\mathcal{T}}$ is an anti-linear operator it follows that the complex modal amplitudes must satisfy $\left[ \left( a_n^+ \right)^* \right] = -\mathbf{S} \cdot \left[ \left( a_n^- \right)^* \right]$ where the "*" superscript stands for complex conjugation. The leading minus sign is due to the property $\tilde{\mathcal{T}}^2 = -\mathbf{1}$. Therefore under these conditions it follows that $\mathbf{S}^{-1} = -\mathbf{S}^*$. For a lossless junction the scattering matrix must satisfy $\mathbf{S}^{\dagger} \cdot \mathbf{S} = \mathbf{1}$ to ensure that the power transported by the scattered waves is identical to the power transported by the incident waves: $\sum_n |a_n^-|^2 = \sum_n |a_n^+|^2$. Therefore, for a lossless $\tilde{\mathcal{T}}$-invariant structure the scattering matrix must be anti-symmetric:

$$\mathbf{S} = -\mathbf{S}^T. \tag{5}$$

In particular, this property implies that the matrices $\mathbf{S}_{11}$ and $\mathbf{S}_{22}$ – which characterize the reflected waves in each waveguide – are also anti-symmetric. The scattering anomaly occurs when $N$ – the number of propagating modes in each waveguide – is odd. In this case, the determinant of $\mathbf{S}_{11}$ and $\mathbf{S}_{22}$ must vanish because the anti-symmetry implies that $\det(\mathbf{S}_{11}) = (-1)^N \det(\mathbf{S}_{11})$. Then, it is possible to choose the incident wave, let us say in waveguide 1, in such a manner that the reflected wave in the same waveguide vanishes. Thus, due to the conservation of the energy, the incoming wave in waveguide 1 must be totally transmitted across the $\tilde{\mathcal{T}}$-invariant lossless junction – independent of its form, shape, or specific material composition – to the waveguide 2. We note in passing, that the same derivation gives a symmetric scattering matrix, $\mathbf{S} = \mathbf{S}^T$, when the operator $\tilde{\mathcal{T}}$ is taken identical to the standard time-reversal operator ($\mathcal{T}^2 = \mathbf{1}$). In this case there is no scattering anomaly, because independent of the value of $N$ the null space of the matrices $\mathbf{S}_{11}$ and $\mathbf{S}_{22}$ is generally trivial. The property $\mathbf{S} = \mathbf{S}^T$ is generally valid for reciprocal electromagnetic networks even in the case of material absorption [33]. The difference in the symmetry of the scattering matrix puts into perspective how the sign of $\tilde{\mathcal{T}}^2$ influences in a decisive manner the wave phenomena.

It is important to underline that the result $\mathbf{S} = -\mathbf{S}^T$ requires that all the possible propagation channels are included in the scattering matrix so that the junction can be considered lossless. For example, if new propagations channels are created when interfacing two different waveguides then these must also be included in the scattering matrix formulation. This may happen when the input and output waveguides are formed by different materials. For simplicity, in this article we restrict our attention to the case where the input and output waveguides are identical.

## B. *Construction of the $\tilde{\mathcal{T}}$-operator*

To construct an operator $\tilde{\mathcal{T}}$ with the desired properties, first we consider a generalized duality transformation of the form $\mathcal{D} = \begin{pmatrix} d_{11}\mathbf{1}_{3\times 3} & d_{12}\mathbf{1}_{3\times 3} \\ d_{21}\mathbf{1}_{3\times 3} & d_{22}\mathbf{1}_{3\times 3} \end{pmatrix}$ that maps the electromagnetic fields as $\mathbf{f} \to \mathcal{D} \cdot \mathbf{f}$ without affecting the space and the time coordinates [32]. The properties of duality transformations are reviewed in Appendix A. We are interested in mappings such that $\mathcal{D}^2 = -\mathbf{1}$, and hence we restrict our attention to duality transformations of the form:

$$\mathcal{D} = \alpha_1 \boldsymbol{\sigma}_x + i\alpha_2 \boldsymbol{\sigma}_y + \alpha_3 \boldsymbol{\sigma}_z, \quad \text{with} \quad \alpha_1^2 - \alpha_2^2 + \alpha_3^2 = -1. \tag{6}$$

The constant coefficients $\alpha_i$ are required to be real-valued for reasons that will be explained shortly. Here, $\boldsymbol{\sigma}_x = \begin{pmatrix} 0 & \mathbf{1}_{3\times 3} \\ \mathbf{1}_{3\times 3} & 0 \end{pmatrix}$, $\boldsymbol{\sigma}_y = \begin{pmatrix} 0 & -i\mathbf{1}_{3\times 3} \\ i\mathbf{1}_{3\times 3} & 0 \end{pmatrix}$, and $\boldsymbol{\sigma}_z$ (defined as before) are 6×6 matrices with the same algebraic properties as the Pauli matrices. This subclass of duality transformations preserves the structure of the Maxwell's equations (1) provided the $\mathbf{g}$ fields are transformed as $\mathbf{g} \to -\mathcal{D}^T \cdot \mathbf{g}$. Note that the considered duality transformations have zero trace ($d_{11} = -d_{22}$) and determinant $\det \mathcal{D} = D^3 = 1$ where $D = d_{11}d_{22} - d_{12}d_{21} = 1$.

The composition of the time-reversal operator $\mathcal{T}$ and the duality transformation $\mathcal{D}$ does *not* give an operator $\tilde{\mathcal{T}}$ with the desired properties. The difficulty is that the operators $\mathcal{T}$ and $\mathcal{D}$ do not commute and because of this $(\mathcal{T} \cdot \mathcal{D})^2 \neq -\mathbf{1}$. To fix this problem we need to introduce a third operator, specifically, a parity (inversion) operator that flips the *z*-spatial coordinate $(x,y,z) \to (x,y,-z)$, i.e., the coordinate perpendicular to the propagation plane (*xoy* plane). A parity operation transforms the electromagnetic fields as $\mathbf{f} \to \mathcal{P} \cdot \mathbf{f}$ and $\mathbf{g} \to \mathcal{P} \cdot \mathbf{g}$ with $\mathcal{P} = \begin{pmatrix} -\mathbf{V} & 0 \\ 0 & \mathbf{V} \end{pmatrix}$ where $\mathbf{V}$ is a 3×3 diagonal matrix with diagonal entries $V_{11} = V_{22} = -V_{33} = 1$. The parity operator leaves the in-plane (*x* and *y* components of the) Poynting vector unchanged. Evidently, $\mathcal{P}$ and $\mathcal{T}$ commute and $\mathcal{P}^2 = \mathbf{1}$. Moreover, $\mathcal{P} \cdot \mathcal{T}$ commutes with the duality transformation $\mathcal{D}$ provided the coefficients $\alpha_i$ are real-valued. Therefore, the operator $\tilde{\mathcal{T}}$ can be defined as the composition of the time-reversal, parity and duality transformations:

$$\tilde{\mathcal{T}} = \mathcal{P} \cdot \mathcal{T} \cdot \mathcal{D} \qquad (7)$$

In summary, it was demonstrated that the above defined operator satisfies $\tilde{\mathcal{T}}^2 = -\mathbf{1}$, and transforms the electromagnetic fields as $\mathbf{f}(\mathbf{r}) \to \tilde{\mathbf{f}}(\mathbf{r}) \equiv \tilde{\mathcal{T}} \cdot \mathbf{f}(\mathbf{V} \cdot \mathbf{r})$ and $\mathbf{g}(\mathbf{r}) \to \tilde{\mathbf{g}}(\mathbf{r}) \equiv \mathcal{P} \cdot \mathcal{T} \cdot (-\mathcal{D}^T) \cdot \mathbf{g}(\mathbf{V} \cdot \mathbf{r})$. As desired, the in-plane components of the Poynting vector are flipped by this transformation.

## C.   $\tilde{\mathcal{T}}$ -invariant photonic systems

A photonic system is invariant under $\tilde{\mathcal{T}}$ when the material matrix satisfies:

$$\mathbf{M}(x,y,z,\omega) = \left[-\mathcal{P}\cdot\mathcal{T}\cdot\mathcal{D}^T\right]\cdot\mathbf{M}(x,y,-z,\omega)\cdot\left[\mathcal{P}\cdot\mathcal{T}\cdot\mathcal{D}\right]^{-1}. \tag{8}$$

This condition ensures that the $\tilde{\mathcal{T}}$ -transformed fields ($\tilde{\mathbf{f}}$ and $\tilde{\mathbf{g}}$) satisfy the Maxwell's equations in the same material structure as the original fields ($\mathbf{f}$ and $\mathbf{g}$). In order that the structure is lossless it is also required that the material matrix is Hermitian symmetric $\mathbf{M} = \mathbf{M}^\dagger$ [32]. Hence, using $\mathbf{M}^* = \mathbf{M}^T$ and $\mathcal{D}^{-1} = -\mathcal{D}$ it is readily found that these properties imply that:

$$\begin{pmatrix} \varepsilon_0\overline{\varepsilon} & \frac{1}{c}\overline{\xi} \\ \frac{1}{c}\overline{\zeta} & \mu_0\overline{\mu} \end{pmatrix}_{(x,y,z)} = \mathcal{D}^T \cdot \begin{pmatrix} \mathbf{V}\cdot\varepsilon_0\overline{\varepsilon}^{-T}\cdot\mathbf{V} & \frac{1}{c}\mathbf{V}\cdot\overline{\zeta}^T\cdot\mathbf{V} \\ \frac{1}{c}\mathbf{V}\cdot\overline{\xi}^T\cdot\mathbf{V} & \mathbf{V}\cdot\mu_0\overline{\mu}^T\cdot\mathbf{V} \end{pmatrix}_{(x,y,-z)} \cdot \mathcal{D} \tag{9}$$

For a fixed $\mathcal{D}$ defined as in Eq. (6), a lossless system is $\tilde{\mathcal{T}}$ -invariant if and only if it satisfies the above equation. In general, $\tilde{\mathcal{T}}$ -invariant systems may include nonreciprocal media because $\tilde{\mathcal{T}} \neq \mathcal{T}$. For simplicity, this article deals only with duality transformations of the form

$$\mathcal{D} = \begin{pmatrix} 0 & A\eta_0 \mathbf{1}_{3\times 3} \\ -A^{-1}\eta_0^{-1}\mathbf{1}_{3\times 3} & 0 \end{pmatrix}, \tag{10}$$

where $\eta_0 = \sqrt{\mu_0/\varepsilon_0}$ is the free-space impedance and $A$ is an arbitrary nonzero dimensionless real parameter. In this case, the $\tilde{\mathcal{T}}$ -invariance reduces to the two simple conditions:

$$\overline{\varepsilon}(x,y,z) = \frac{1}{A^2}\mathbf{V}\cdot\overline{\mu}^T(x,y,-z)\cdot\mathbf{V}, \qquad \overline{\xi}(x,y,z) = -\mathbf{V}\cdot\overline{\xi}^T(x,y,-z)\cdot\mathbf{V}. \tag{11}$$

It is implicitly assumed that $\mathbf{M} = \mathbf{M}^\dagger$ so that all the materials are lossless, and in particular $\overline{\overline{\zeta}} = \overline{\overline{\xi}}^\dagger$. It is important to highlight that these conditions need to be satisfied only at the frequency of interest, i.e., they are not required to hold for all frequencies. Moreover, except for the special $z=0$ symmetry plane, the conditions (11) link the material parameters at *different* points of space. This property turns out to be quite important because it gives us increased design flexibility. Finally, it is underlined that different from most works on topological photonics the constraints (11) apply as well to generic 3D-dimensional structures, and hence offer the possibility to a design a 3D waveguide insensitive to backscattering. This idea will be developed ahead.

At this point it is worth noting that the pseudo-time reversal operator $\mathcal{T}_p = \mathcal{T} \cdot \boldsymbol{\sigma}_x$ introduced in Ref. [24] is not equivalent to $\tilde{\mathcal{T}} = \mathcal{P} \cdot \mathcal{T} \cdot \mathcal{D}$. Indeed, while the latter operator acts on the spatial coordinates through the action of the parity operator, the former does not. For two-dimensional problems with $\partial / \partial z = 0$ the operator $\tilde{\mathcal{T}}$ reduces to [taking $A\eta_0 = -1$ in Eq. (10)] $\tilde{\mathcal{T}} = \mathcal{T} \cdot \begin{pmatrix} 0 & \mathbf{V} \\ \mathbf{V} & 0 \end{pmatrix}$ which has a form alike but not equivalent to $\mathcal{T}_p = \mathcal{T} \cdot \boldsymbol{\sigma}_x$.

It is also important to highlight that for standard isotropic materials the $\tilde{\mathcal{T}}$-invariance reduces to $\varepsilon(x, y, z) = \frac{1}{A^2} \mu(x, y, -z)$. This condition coincides precisely with that derived in Ref. [26] to ensure symmetry protected light transport in bulk waveguides. Our theory extends the result of Ref. [26] to general anisotropic and bianisotropic photonic platforms.

## D. $\tilde{\mathcal{T}}$-invariant materials

In systems uniform along the *z*-direction the constraints (11) must be individually satisfied by all the materials in the system, or in different words all the media need to be $\tilde{\mathcal{T}}$-invariant. The parameters of a $\tilde{\mathcal{T}}$-invariant lossless material are linked by:

$$\overline{\varepsilon} = \frac{1}{A^2}\mathbf{V}\cdot\overline{\mu}^T\cdot\mathbf{V}, \qquad \overline{\xi} = -\mathbf{V}\cdot\overline{\xi}^T\cdot\mathbf{V}, \qquad (\tilde{\mathcal{T}}\text{-invariant material}). \qquad (12)$$

A magneto-electric tensor compatible with the above conditions is necessarily of the form:

$$\overline{\xi} = \xi_{yx}\,\hat{\mathbf{z}}\times\mathbf{1}_{3\times 3} + \left[\xi_{xz}\left(\hat{\mathbf{x}}\otimes\hat{\mathbf{z}}+\hat{\mathbf{z}}\otimes\hat{\mathbf{x}}\right)+\xi_{yz}\left(\hat{\mathbf{y}}\otimes\hat{\mathbf{z}}+\hat{\mathbf{z}}\otimes\hat{\mathbf{y}}\right)\right], \qquad (13)$$

with $\xi_{yx},\xi_{xz},\xi_{yz}$ arbitrary complex-valued numbers. The first term determines an anti-symmetric type coupling and the second term a symmetric type coupling.

## III. Scattering anomaly with Gyrotropic Media

To begin with, we restrict our attention to structures uniform along the *z*-direction and consider wave propagation with $\partial/\partial z = 0$ (two-dimensional problem). The simplest example of a "scattering anomaly" occurs in a system with no magneto-electric response ($\overline{\xi}=\overline{\zeta}=0$). In this case, the $\tilde{\mathcal{T}}$-invariance [Eq. (12)] reduces simply to $\overline{\varepsilon} = \frac{1}{A^2}\mathbf{V}\cdot\overline{\mu}^T\cdot\mathbf{V}$ for some fixed scaling parameter *A*. As illustrated in Fig. 2d, a two-dimensional waveguide with this symmetry can be realized by pairing two homogeneous materials. The material interface is at the plane $y=0$ so that the propagation is along the *x*-direction. The material in the region $y>0$ is assumed to have a gyrotropic response described by (for simplicity we fix *A*=1 from here on):

$$\bar{\bar{\varepsilon}} = \begin{pmatrix} \varepsilon_{11} & \varepsilon_{12} & 0 \\ -\varepsilon_{12} & \varepsilon_{22} & 0 \\ 0 & 0 & \varepsilon_{33} \end{pmatrix}, \qquad \bar{\bar{\mu}} = \bar{\bar{\varepsilon}}^{-T}, \tag{14a}$$

$$\varepsilon_{11} = \varepsilon_{22} = 1 + \frac{\omega_p^2}{\omega_0^2 - \omega^2}, \qquad \varepsilon_{12} = -\varepsilon_{21} = \frac{-i\omega_g \omega}{\omega_0^2 - \omega^2}, \qquad \varepsilon_{33} = 1. \tag{14b}$$

Here, $\omega_0$ is the resonance frequency, $\omega_p$ determines the resonance strength, and $\omega_g$ determines the nonreciprocal response strength; the three parameters must satisfy $\omega_p \geq \sqrt{\omega_0 |\omega_g|}$. The proposed gyrotropic material is nonreciprocal: it has an electric response analogous (but not equivalent) to that of an electric-plasma biased with a static magnetic field [34], and a magnetic response equivalent (with a suitable choice of the relevant parameters) to that of a lossless ferrite biased with a static magnetic field [33]. The parameter $\omega_g$ can be either positive or negative, depending on the orientation of the biasing field. The material in the region $y < 0$ is taken as an anisotropic-type dielectric with:

$$\bar{\bar{\varepsilon}} = \bar{\bar{\mu}} = \varepsilon_{a,\parallel}\left(\hat{\mathbf{x}}\hat{\mathbf{x}} + \hat{\mathbf{y}}\hat{\mathbf{y}}\right) + \varepsilon_{a,zz}\hat{\mathbf{z}}\hat{\mathbf{z}}, \qquad \text{with } \varepsilon_{a,\parallel} = 1 + \frac{\omega_{e,a}^2}{\omega_{0,a}^2 - \omega^2} \text{ and } \varepsilon_{a,zz} = 1. \tag{15}$$

In the proposed waveguide the condition (11) holds for every frequency, but it is underlined that the $\tilde{\mathcal{T}}$-invariance only needs to be satisfied in a narrowband frequency region. Indeed, the key feature of the scattering anomaly is that the scattering matrix is anti-symmetric and the number of modes $N$ is odd at the frequency of interest.

Figures 2a and 2b represent the band diagrams of the two bulk materials for in-plane propagation. The band structure of the gyrotropic material is calculated as in Ref. [12]. Note that the band structure is independent of the direction of propagation in the $xoy$ plane. Because of the $\tilde{\mathcal{T}}$-invariance all the bands are doubly degenerate and the $s$- polarized waves (with $E_z \neq 0$ and $H_z = 0$) and the *p*-polarized waves (with $E_z = 0$ and $H_z \neq 0$) are completely decoupled. The insets of the figures indicate the Chern numbers for each band subset and for each polarization ($\mathcal{C}_s$ and $\mathcal{C}_p$ for *s* and *p* polarizations, respectively). The Chern numbers are computed using a high-frequency spatial cut-off in the material response, as detailed in Refs. [12, 13]. First of all, we note that for all bands $\mathcal{C}_s + \mathcal{C}_p = 0$. This property is a consequence of the invariance of the material response under the $\tilde{\mathcal{T}}$ transformation [Eq. (12)], which can be shown to imply that the Berry curvature is an odd function of the wave vector, similar to the electronic case. Note that $\tilde{\mathcal{T}}$ maps the eigenmodes as $\mathbf{f}_{n\mathbf{k}} \to \tilde{\mathcal{T}} \cdot \mathbf{f}_{n\mathbf{k}}$, such that the frequency and wave vector are transformed as $(\omega_{n\mathbf{k}}, \mathbf{k}) \to (\omega_{n\mathbf{k}}, -\mathbf{k})$. As expected, the Chern numbers of the anisotropic dielectric are trivial. In contrast, the Chern numbers of the gyrotropic material are nonzero for a fixed polarization. The property $\mathcal{C}_p = -\mathcal{C}_s$ can be intuitively understood as a consequence of $\bar{\bar{\mu}} = \bar{\bar{\varepsilon}}^{-T}$, which implies that the nonreciprocal coupling of the electric and magnetic responses is opposite for the *s* and *p* polarizations (note that the matrix transposition swaps the sign of the nondiagonal elements).

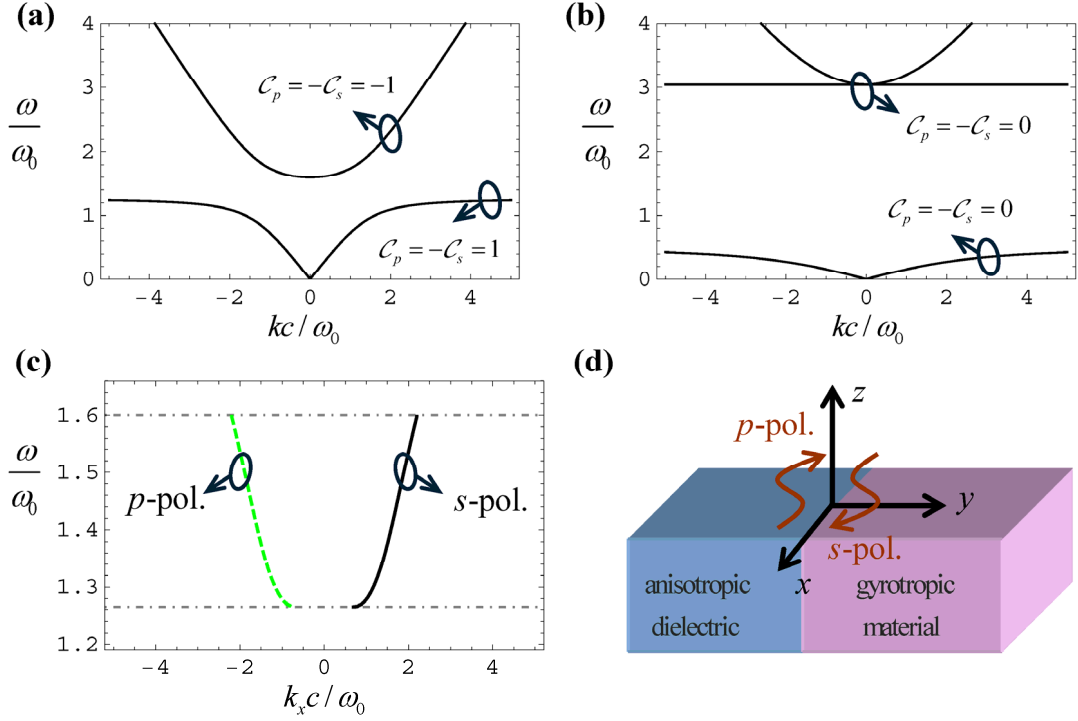

Fig. 2. (Color online) **Scattering anomaly with gyrotropic media.** (a) Band diagram $\omega$ vs. $k$ for a $\tilde{\mathcal{T}}$-invariant gyrotropic material with $\omega_g = 0.6\omega_0$ and $\omega_p = \sqrt{|\omega_g|\omega_0}$. The insets indicate the Chern numbers for each band and for each polarization. (b) Similar to (a) but for a $\tilde{\mathcal{T}}$-invariant anisotropic dielectric with $\omega_{0,a} = 0.5\omega_0$ and $\omega_{e,a} = 3\omega_0$. (c) Dispersion of the guided modes supported by an interface of the two materials. The common bandgap is delimited by the two dot-dashed horizontal gray lines. (d) Sketch of the waveguide geometry. In panels (a) and (b) all the bands are doubly degenerate. In panel (c) the guided modes are *not* degenerate, and there is a single mode propagating along the positive $x$-axis, corresponding to a scattering anomaly with $N=1$.

It is interesting to point out that in the same manner as the Kane-Mele model corresponds to two copies of the Haldane model [35], our system corresponds as well to a duplication of the response of a standard gyrotropic material (e.g., a duplication of the response to *s*-polarized waves of a biased ferrite). The two uncoupled polarizations are linked by the $\tilde{\mathcal{T}}$ operator. Similar to the theory of electronic topological insulators [29, 30, 31, 35], when the $\tilde{\mathcal{T}}$-invariance holds for all frequencies, as in the present example, it is possible to introduce a $\mathbb{Z}_2$ topological index defined by $\Delta = \left[ \left( \mathcal{C}_p - \mathcal{C}_s \right) / 2 \right] \mod 2$. Clearly, with this classification the $\tilde{\mathcal{T}}$-invariant gyrotropic material is topologically nontrivial because for the lower band subset (below the bandgap) one has $\Delta = 1$. Totally different, the anisotropic dielectric is topologically trivial ($\Delta = 0$). Even though this topological classification is illuminating, it should be underlined that the calculation of topological invariants depends on the *global* properties of the band structure, whereas our general scattering anomaly condition [Eq. (9)] is intrinsically a *single frequency* property and thus relies on simpler assumptions easier to test and to meet. Moreover, it does not require a detailed knowledge of the material dispersion and it is not necessary that the material is lossless or $\tilde{\mathcal{T}}$-invariant for *all* frequencies below the band gap.

The dispersion of the edge modes for *p*-polarized waves is determined by the equation [12, 13, 36]:

$$\frac{\gamma_a}{\varepsilon_a} + \frac{\gamma_g}{\varepsilon_{ef}} = \frac{1}{\varepsilon_{ef}} \frac{\varepsilon_{12} i k_x}{\varepsilon_{11}}, \qquad (p\text{-pol.}). \qquad (16)$$

Here, $k_x$ is the edge state propagation constant, $\varepsilon_{11}, \varepsilon_{12}$ describe the response of the gyrotropic material [Eq. (14)], $\varepsilon_{ef} = \frac{\varepsilon_{11}^2 + \varepsilon_{12}^2}{\varepsilon_{11}}$, $\gamma_g = \sqrt{k_x^2 - \left( \omega / c \right)^2 \varepsilon_{ef}}$ and $\gamma_a = \sqrt{k_x^2 - \left( \omega / c \right)^2 \varepsilon_a}$. The dispersion equation for *s*-polarized waves is given by a similar expression with the symbol "$\varepsilon$" replaced by the symbol "$\mu$" in all the formulas. When $\overline{\overline{\varepsilon}} = \overline{\overline{\mu}}^T$ the dispersion equation of the *s*-polarized waves can be simply written as

$$\frac{\gamma_a}{\varepsilon_a}+\frac{\gamma_g}{\varepsilon_{ef}}=-\frac{1}{\varepsilon_{ef}}\frac{\varepsilon_{12}ik_x}{\varepsilon_{11}}$$ (notice the change of the leading sign in the right-hand side term).

Figure 2c depicts the calculated edge state dispersion for the system under study. Crucially, in the intersection of the material bandgaps there is a *single* edge mode ($N=1$) propagating along the +$x$-direction. Thus, this system provides a remarkable example of a scattering anomaly in optics. Because of the $\tilde{\mathcal{T}}$-invariance there is also a single edge mode propagating in the opposite direction. The two modes have distinct polarization states and are evidently uncoupled. In this example, *p*-polarized waves propagate exclusively towards the –$x$-direction, while *s*-polarized waves propagate exclusively towards the +$x$-direction (see Fig. 2d). Thus, provided the $\tilde{\mathcal{T}}$-invariance is unbroken, that the materials are lossless and that the two edge modes remain the only allowed propagation channels, the edge waveguide can be deformed at will, and still the energy will flow along the edges with no backscattering.

To illustrate this consider a sinuous and meandering deformation of the waveguide, as shown in Fig. 3. The fields radiated by an emitter placed in the vicinity of the interface in the gyrotropic region were calculated using a full wave commercial electromagnetic simulator [37]. For simplicity in the numerical simulation it is assumed that the anisotropic dielectric is an opaque material with permittivity and permeability tensors as in Eq. (15) and $\varepsilon_{a,\parallel}=\mu_{a,\parallel}\to-\infty$. The edge states supported by a planar interface of this opaque anisotropic dielectric and the original gyrotropic material have a dispersion $\omega$ vs $k_x$ completely analogous to that of the example of Fig. 2c. In particular, at the frequency $\omega=1.4\omega_0$ – which lies in the bandgap of both materials – the edge modes have the guided wave number $k_x=\pm0.86\omega_0/c$.

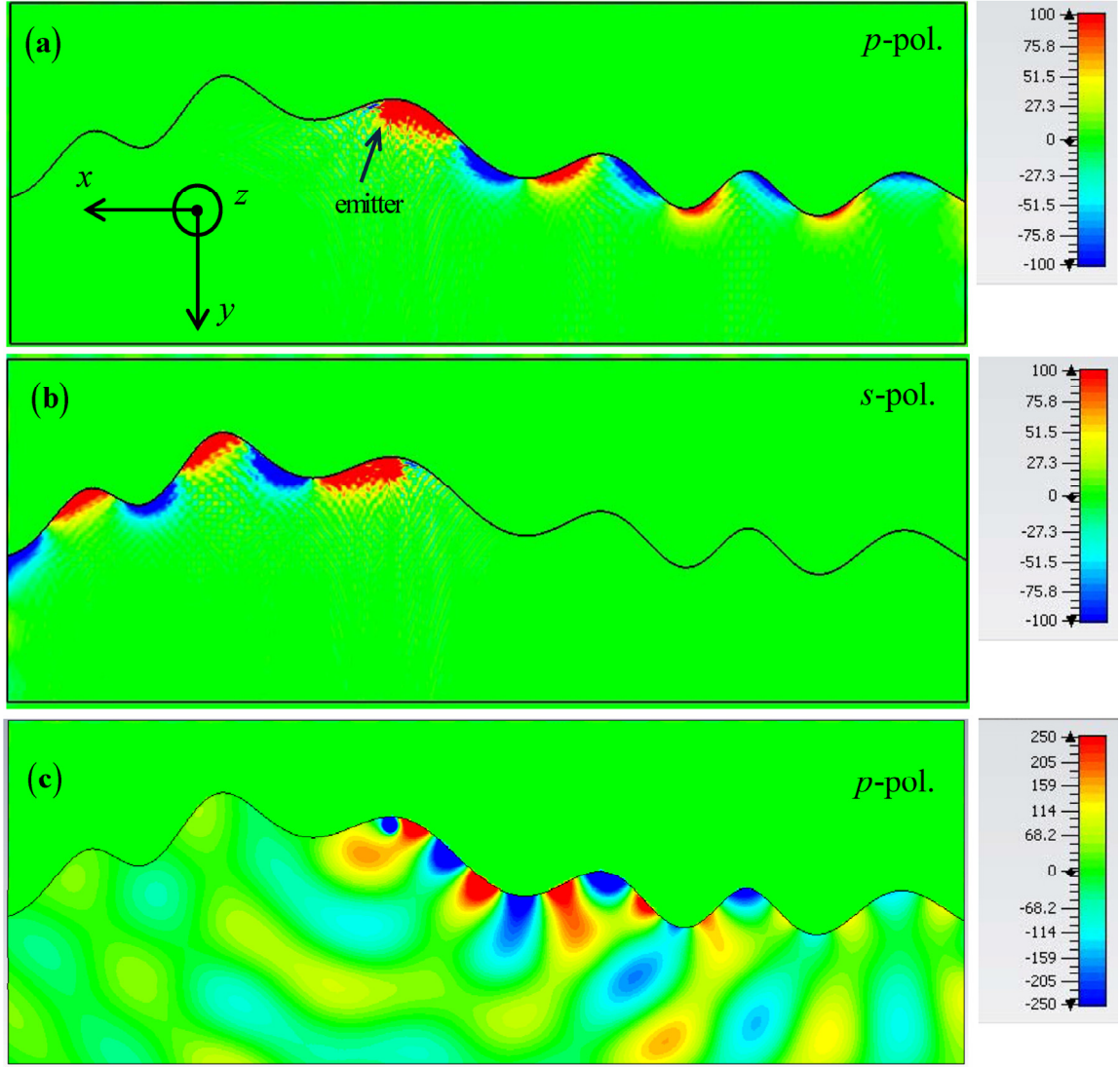

Fig. 3. (Color online) **Waveguiding immune to backscattering in a $\tilde{\mathcal{T}}$-invariant meandering structure.** The gyrotropic material (below the interface) has the same parameters as in Fig. 2, and the anisotropic dielectric (above the interface) is an opaque material characterized by $\varepsilon_{a,\|} = \mu_{a,\|} \to -\infty$ and $\varepsilon_{a,zz} = \mu_{a,zz} = 1$. The width of the plot region is $9.3\lambda_0$ (free-space wavelengths) at $\omega = 1.4\omega_0$. **(a)** Snapshot in time of the *p*-polarized field ($H_z$) emitted by a magnetic line source at $\omega = 1.4\omega_0$. **(b)** Snapshot in time of the *s*-polarized field ($E_z$) emitted by an electric line source at $\omega = 1.4\omega_0$. **(c)** Similar to (a) but for $\omega = 1.8\omega_0$.

Figures 3a and 3b show time snapshots of the fields emitted by a magnetic line source and by an electric line source, respectively. A magnetic (electric) line emitter excites only *p*- (*s*-) polarized waves in the waveguide. Clearly, the radiated waves propagate with no backscattering along the edge, being totally insensitive to the effect of bending. Indeed, due to the scattering anomaly the edge states that propagate in opposite directions are totally decoupled. Moreover, consistent with the analytical results the *p*-polarized waves propagate towards the *–x* direction, whereas the *s*-polarized waves propagate towards the *+x*-direction. The field animations of the emitted fields can be found in the supplemental materials [38].

Quite differently, when the oscillation frequency is outside the bandgap of the gyrotropic material (Fig. 3c) a continuum of propagation channels becomes available in the bulk gyrotropic material. In this case the emitted wave consists of both space and surface waves. Distinct from Figs. 3a and 3b, the surface wave is highly sensitive to the bending effect because the deformation of the waveguide leads to radiation leakage due to the coupling with the radiation continuum.

One may argue that the role of the $\tilde{\mathcal{T}}$-invariance in this system is a bit artificial because other than ensuring that the counter-propagating edge modes have identical dispersions the $\tilde{\mathcal{T}}$-invariance does not play a relevant role. Indeed, the bidirectional backscattering-free waveguiding can evidently be achieved in similar material platforms even when $\bar{\bar{\varepsilon}} \neq \bar{\bar{\mu}}^T$ due to the intrinsic topological properties of the electric and magnetic responses taken separately. It will be shown in the next section that $\tilde{\mathcal{T}}$-invariant gyrotropic materials have other unique properties that make, indeed, this class of materials rather special.

It is relevant to discuss the microscopic mechanisms that may enable a dual gyrotropic response with $\overline{\varepsilon} = \overline{\mu}^T$ at some frequency of operation using a biasing magnetic field $B_0$. The gyrotropic magnetic response is typically due to the precession of magnetic dipoles around some axis determined by the biasing field [33]. On the other hand, the gyrotropic electric response may be due to the action of the magnetic (Lorentz) force on the free carriers of a plasma [34]. The interesting point is that for typical material dispersions the sign of $i\varepsilon_{12}$ ($i\mu_{12}$) depends only on the sign of the cyclotron frequency $\omega_0 = qB_0/m_0$, where $q$ is the electric charge of the current carriers and $m_0$ is the corresponding mass. In particular, when all the microscopic currents are due to the motion of electrons, it follows that to have $\varepsilon_{12} = -\mu_{12}$ it is necessary that the charge carriers that determine the electric response feel a biasing field symmetric to that felt by the currents loops (magnetic dipoles) that create the magnetic response. In this case, it seems that the biasing field is required to vary in space at the microscopic level. Alternatively, one can imagine that the electric gyrotropic response is due to the motion of positively charged particles (ions or holes) whereas the gyrotropic magnetic response is due to the motion of electrons (current loops) with negative charge. In this situation, it is in principle possible to ensure that $\overline{\varepsilon} = \overline{\mu}^T$ using a constant biasing magnetic field. Any of the outlined mechanisms is evidently rather complex from a practical standpoint. Notably, it will be proven in the next section that there is a relatively simple way to mimic the electromagnetic response of $\tilde{\mathcal{T}}$-invariant gyrotropic materials without a biasing magnetic field.

# IV. Scattering anomaly with Reciprocal Media

Next, it is demonstrated that it is possible to have a scattering anomaly even when all the involved (lossless) media are reciprocal, i.e., invariant under the usual time-reversal operation $\mathcal{T}$.

At first sight, the invariance of an optical system under both $\tilde{\mathcal{T}}$ and $\mathcal{T}$ may appear to lead to a contraction. Indeed, as demonstrated in Sect. II the invariance of a system under the operator $\tilde{\mathcal{T}}$ implies that the scattering matrix satisfies $\mathbf{S} = -\mathbf{S}^T$. On the other hand, the invariance under the time reversal operator $\mathcal{T}$ implies that $\mathbf{S} = +\mathbf{S}^T$. This may suggest that when a system is simultaneously invariant under $\mathcal{T}$ and $\tilde{\mathcal{T}}$ one obtains the paradoxical result $\mathbf{S} = \mathbf{0}$, which is evidently inconsistent with the conservation of energy in the lossless junction. This might be interpreted as an indication that the invariance under $\mathcal{T}$ is incompatible with the invariance under $\tilde{\mathcal{T}}$. However, that conclusion is wrong.

The flaw in the reasoning is that the definition of the scattering matrix *depends* on the basis of modes used to expand the fields. Indeed, similar to Sect. II let us suppose that the incident wave is expanded in terms of some $\mathbf{f}_n^+$ modes. The property $\mathbf{S} = -\mathbf{S}^T$ requires the $\tilde{\mathcal{T}}$-invariance of the system and assumes that the scattered wave is expanded in terms of the basis $\tilde{\mathcal{T}} \cdot \mathbf{f}_n^+$ (see Sect. II). Similarly, the property $\mathbf{S} = +\mathbf{S}^T$ relies on $\mathcal{T}$-invariance of the system and assumes that the scattered wave is expanded in terms of the basis $\mathcal{T} \cdot \mathbf{f}_n^+$. Evidently, the two bases, $\tilde{\mathcal{T}} \cdot \mathbf{f}_n^+$ and $\mathcal{T} \cdot \mathbf{f}_n^+$, are different and hence the scattering matrix that satisfies $\mathbf{S} = -\mathbf{S}^T$ is *not* the same as the scattering matrix that satisfies $\mathbf{S} = +\mathbf{S}^T$. Note that both scattering matrices describe equally well the relevant wave phenomena, but they are computed using two different bases of outgoing modes. In summary, this discussion shows that the symmetries $\mathbf{S} = -\mathbf{S}^T$ and $\mathbf{S} = +\mathbf{S}^T$ are compatible and that the invariance of a system under both $\mathcal{T}$ and $\tilde{\mathcal{T}}$ does not lead to inconsistencies. In the following, we present an example of reciprocal system invariant under $\tilde{\mathcal{T}}$.

## A. $\tilde{\mathcal{T}}$-invariant reciprocal materials

To begin with, we consider the subclass of reciprocal media that satisfies Eq. (12). For lossless reciprocal media the permittivity and permeability tensors are required to be real-valued and symmetric whereas the magneto-electric coupling tensors $\bar{\zeta} = -\bar{\xi}^T$ are required to be pure imaginary [28, 32]. Thus, from Eq. (12) the permittivity and permeability of $\tilde{\mathcal{T}}$-invariant materials must satisfy:

$$\bar{\varepsilon} = \frac{1}{A^2} \mathbf{V} \cdot \bar{\mu} \cdot \mathbf{V} . \tag{17a}$$

Moreover, since for reciprocal media $\xi_{yx}, \xi_{xz}, \xi_{yz}$ are pure imaginary, the magneto-electric coupling tensor (13) can be written as:

$$\bar{\xi} = -i\Omega \hat{\mathbf{z}} \times \mathbf{1}_{3\times 3} + \left[ -i\chi_{xz} \left( \hat{\mathbf{x}} \otimes \hat{\mathbf{z}} + \hat{\mathbf{z}} \otimes \hat{\mathbf{x}} \right) - i\chi_{yz} \left( \hat{\mathbf{y}} \otimes \hat{\mathbf{z}} + \hat{\mathbf{z}} \otimes \hat{\mathbf{y}} \right) \right], \tag{17b}$$

with $\Omega, \chi_{xz}, \chi_{yz}$ real-valued. Materials with a traceless magneto-electric coupling were originally proposed by Saadoun and Engheta, and generically can be realized as metamaterials formed by $\Omega$-shaped inclusions [39, 40]. Following Ref. [32], we refer to a traceless asymmetric coupling – corresponding to the first term in the right-hand side of Eq. (17b) – as an $\Omega$-type coupling. On the other hand, the second term in Eq. (17b) determines a different type of magneto-electric response that to the best of our knowledge was not previously discussed in the context of topological materials. We refer to this type of traceless symmetric magneto-electric coupling as "pseudochiral" coupling [32]. In this article, we focus our attention in the $\Omega$-type coupling.

Notably, the class of $\tilde{\mathcal{T}}$-invariant systems includes as a particular case the topological photonic insulators discovered in Ref. [18] using an analogy with the Kane-Mele model [35]. Specifically, the structures studied in Ref. [18] turn out to be photonic crystals formed by $\tilde{\mathcal{T}}$-invariant $\Omega$-materials, wherein the operator $\mathbf{V}$ plays no role and the material parameters are independent of *z*. Importantly, the crucial point that a material response consistent with Eq. (12) implies that the scattering matrix is anti-symmetric was missed in earlier works. In addition, it is stressed that our general theory applies to three-dimensional photonic platforms, whereas earlier studies on the $\Omega$-medium were restricted to two-dimensional systems.

## B. Scattering anomaly with $\Omega$-media

Let us consider an $\Omega$-type coupling such that the material matrix is of the form:

$$\mathbf{M} = \begin{pmatrix} \varepsilon_0 \overline{\varepsilon} & -i\Omega \frac{1}{c} \hat{\mathbf{z}} \times \mathbf{1}_{3\times 3} \\ -i\Omega \frac{1}{c} \hat{\mathbf{z}} \times \mathbf{1}_{3\times 3} & \mu_0 \overline{\mu} \end{pmatrix}, \qquad (18)$$

with $\overline{\varepsilon} = \overline{\mu} = \varepsilon_\parallel (\hat{\mathbf{x}}\hat{\mathbf{x}} + \hat{\mathbf{y}}\hat{\mathbf{y}}) + \varepsilon_{zz}\hat{\mathbf{z}}\hat{\mathbf{z}}$, $\varepsilon_\parallel = 1 + \frac{\omega_p^2}{\omega_0^2 - \omega^2}$ and $\Omega = \frac{\omega \omega_\Omega}{\omega_0^2 - \omega^2}$. The parameters $\omega_0, \omega_p, \omega_\Omega$ determine the resonance frequency, the electric and the magnetic resonances strength and the $\Omega$-coupling strength, respectively. These parameters are required to satisfy $\omega_p \geq \sqrt{|\omega_\Omega|\omega_0}$ so that the energy stored in the medium is bounded from below [12] ($\partial(\omega\mathbf{M})/\partial\omega$ must be a positive definite matrix), and $\omega_\Omega$ can be either a negative or a positive number. For simplicity, the *zz*-component of the material parameters is assumed constant and identical to the unity: $\varepsilon_{zz} = \mu_{zz} = 1$. Metamaterial designs that may implement the required response (at least at a single frequency) can be found in Ref. [18].

As illustrated in Fig. 4d, we consider a waveguide formed by pairing the anisotropic Ω-medium with an anisotropic dielectric with material dispersion as in Eq. (15). Evidently, both materials are simultaneously reciprocal and $\tilde{\mathcal{T}}$-invariant with $A = 1$. The band structures of the two materials are depicted in Fig. 4a, for the Ω-medium, and in Fig. 4b, for the anisotropic dielectric. It turns out that due to the Ω-coupling the plane waves supported by the bulk Ω-medium do not split into *s* and *p* modes [28, 40]. Yet, it is possible to classify the plane waves as "*s*-type" and "*p*-type" and these have the following dispersions for in-plane (*xoy*) propagation [28, 40]:

$$k^2 = \frac{\omega^2}{c^2}\left(\varepsilon_{zz}\mu_{\parallel} - \frac{\varepsilon_{zz}}{\varepsilon_{\parallel}}\Omega^2\right), \qquad (s\text{-type}), \qquad (19a)$$

$$k^2 = \frac{\omega^2}{c^2}\left(\varepsilon_{\parallel}\mu_{zz} - \frac{\mu_{zz}}{\mu_{\parallel}}\Omega^2\right), \qquad (p\text{-type}). \qquad (19b)$$

Here, $\mu_{\parallel}, \mu_{zz}$ are the in-plane and *zz* components of the permeability of the Ω-medium. Evidently, when the Ω-medium is $\tilde{\mathcal{T}}$-invariant, one has $\varepsilon_{zz} = \mu_{zz}$ and $\varepsilon_{\parallel} = \mu_{\parallel}$ so that the two polarizations are degenerate. Hence, each photonic band in Figs. 4a and 4b is doubly degenerate. The band structure of the Ω-medium is calculated using Eq. (19).

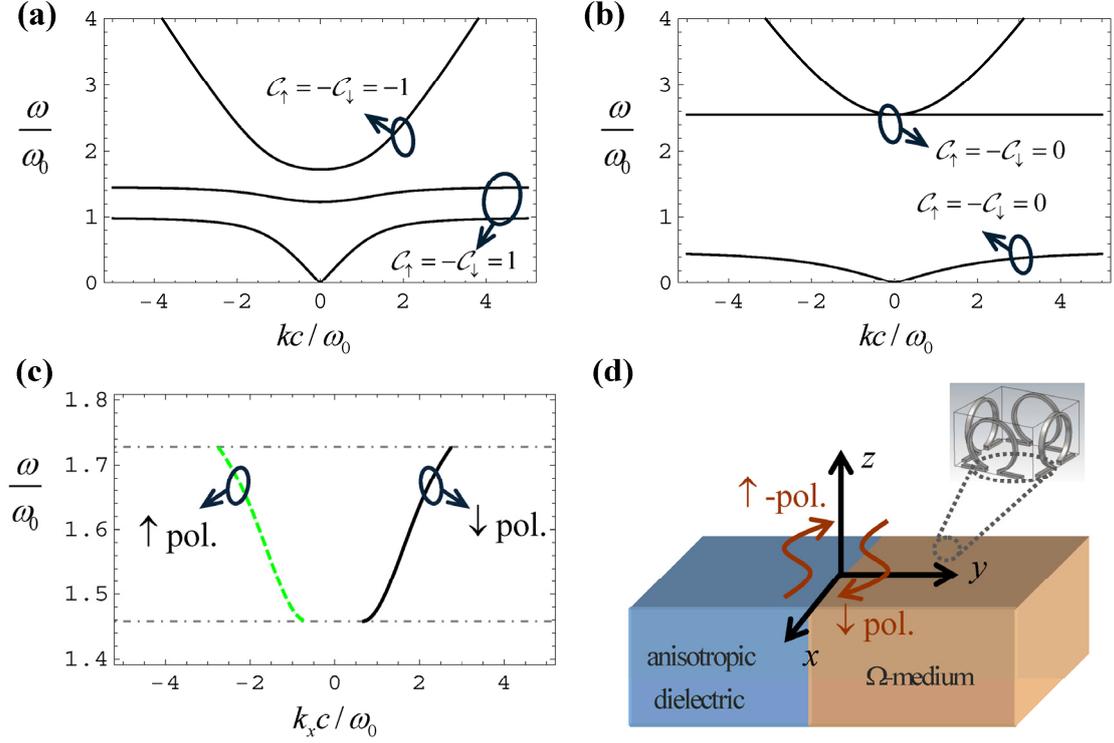

Fig. 4. (Color online) **Scattering anomaly with Ω-media.** (a) Band diagram $\omega$ vs. $k$ for a $\tilde{\mathcal{T}}$-invariant Ω-medium with $\omega_\Omega = 0.5\omega_0$ and $\omega_p = 1.5\sqrt{|\omega_\Omega|\omega_0}$. (b) Similar to (a) but for a $\tilde{\mathcal{T}}$-invariant anisotropic dielectric with $\omega_{0,a} = 0.5\omega_0$ and $\omega_{e,a} = 2.5\omega_0$. (c) Dispersion of the guided modes supported by an interface of the two materials. (d) Sketch of the waveguide geometry.

Detailed calculations show that the dispersion of the edge states supported by the planar waveguide of Fig. 4d is determined by (the derivation is omitted for conciseness):

$$\left(\gamma_{\Omega,s} + \gamma_{a,s}\frac{n_{\Omega,s}^2}{\mu_{a,\parallel}\varepsilon_{zz}}\right)\left(\gamma_{\Omega,p} + \gamma_{a,p}\frac{n_{\Omega,p}^2}{\varepsilon_{a,\parallel}\mu_{zz}}\right) - \frac{\Omega^2}{\varepsilon_\parallel \mu_\parallel}k_x^2 = 0, \qquad (20)$$

where $k_x$ is the propagation constant of the guided mode, the parameters $\varepsilon_\parallel, \mu_\parallel, \varepsilon_{zz}, \mu_{zz}, \Omega$ determine the response of the Ω-medium and $\varepsilon_{a,\parallel}, \mu_{a,\parallel}, \varepsilon_{a,zz}, \mu_{a,zz}$ determine the response of the anisotropic dielectric. In the above, we define $\gamma_{\Omega,i} = \sqrt{k_x^2 - n_{\Omega,i}^2 \omega^2/c^2}$ ($i=s,p$) with

$$n_{\Omega,s}^2 = \varepsilon_{zz}\mu_\| - \frac{\varepsilon_{zz}}{\varepsilon_\|}\Omega^2 \quad \text{and} \quad n_{\Omega,p}^2 = \mu_{zz}\varepsilon_\| - \frac{\mu_{zz}}{\mu_\|}\Omega^2 \quad \text{and} \quad \gamma_{a,i} = \sqrt{k_x^2 - n_{a,i}^2\omega^2/c^2} \quad (i=s,p) \text{ with}$$

$n_{a,s}^2 = \varepsilon_{a,zz}\mu_{a,\|}$ and $n_{a,p}^2 = \mu_{a,zz}\varepsilon_{a,\|}$. The edge modes are usually hybrid modes and cannot be classified as either *s* or *p* polarized.

Interestingly, when both materials are $\tilde{\mathcal{T}}$-invariant and the *zz* components of the permittivity and permeability are identical to the unity the dispersion equation simplifies to:

$$\frac{\gamma_\Omega}{n_\Omega^2} + \frac{\gamma_a}{\varepsilon_{a,\|}} = \pm \frac{1}{n_\Omega^2}\frac{\Omega}{\varepsilon_\|}k_x, \tag{21}$$

with $n_\Omega^2 = \varepsilon_\| - \Omega^2/\varepsilon_\|$, $\gamma_\Omega = \sqrt{k_x^2 - n_\Omega^2\omega^2/c^2}$ and $\gamma_a = \sqrt{k_x^2 - \varepsilon_{a,\|}\omega^2/c^2}$. Analogous to Ref. [18], the edge modes associated with the "+" sign are polarized in such a way that $E_z = +\eta_0 H_z$, whereas edge modes associated with the "−" sign satisfy $E_z = -\eta_0 H_z$. This property assumes that the Ω-material fills the region $y > 0$. A justification of this result will be given in the subsection IV.C. We designate the polarizations $E_z = +\eta_0 H_z$ and $E_z = -\eta_0 H_z$ as spin up (↑) and spin down (↓), respectively. The numerically calculated dispersion of the edge modes is represented in Fig. 4c. As seen, the Ω-medium waveguide supports a single edge state (with spin down polarization) for propagation along the +*x*-direction. Thus, the considered waveguide is characterized by a scattering anomaly (*N=1*). Hence, despite the time reversal invariance of the involved media, it follows from Sect. II that the interface of the two materials can be arbitrarily bent and still the edge modes are totally insensitive to the deformations, analogous to the simulations of Fig. 3. The electromagnetic response of the Ω-medium is not implemented in commercial electromagnetic solvers [37]. Surprisingly, it is proven in the next subsection that the wave propagation in $\tilde{\mathcal{T}}$-invariant Ω-waveguides is intrinsically related to the propagation in $\tilde{\mathcal{T}}$-invariant gyrotropic waveguides, such that the numerical solution of a problem involving the dual-symmetric Ω-medium can be obtained from the numerical solution of a problem involving a dual-symmetric gyrotropic-medium.

## C. Relation with dual-symmetric gyrotropic-materials

So far, we have used duality transformations as a tool to construct an anti-linear operator $\tilde{\mathcal{T}}$ that flips the relevant components of the Poynting vector and satisfies $\tilde{\mathcal{T}}^2 = -\mathbf{1}$. This theory has enabled us to identify photonic platforms (e.g., waveguides formed by an Ω-medium or a gyrotropic medium and an anisotropic dielectric) wherein the wave propagation is immune to backscattering.

Crucially, duality theory can be used to discover alternative photonic systems wherein the wave propagation is also protected against backscattering. Indeed, a generic duality mapping $\mathcal{D} = \begin{pmatrix} d_{11}\mathbf{1}_{3\times 3} & d_{12}\mathbf{1}_{3\times 3} \\ d_{21}\mathbf{1}_{3\times 3} & d_{22}\mathbf{1}_{3\times 3} \end{pmatrix}$ transforms a solution $\mathbf{f} = \mathbf{f}(\mathbf{r})$ of Maxwell's equations in some system characterized by a material matrix $\mathbf{M}(\mathbf{r})$ into another solution of Maxwell's equations $\mathbf{f}' = \mathbf{f}'(\mathbf{r})$ in a structure characterized by a material matrix $\mathbf{M}'(\mathbf{r})$. Here, $\mathcal{D}$ is determined by arbitrary real-valued coefficients $d_{ij}$ independent of the spatial coordinates and such that $D = d_{11}d_{22} - d_{21}d_{12} = 1$ (the matrix $\mathcal{D}$ considered here is totally unrelated to the matrix $\mathcal{D}$ used to define the operator $\tilde{\mathcal{T}}$ and in particular $\mathcal{D}$ is not required to be traceless). The explicit relation between $\mathbf{M}$ and $\mathbf{M}'$ is given by Eq. (A2) of Appendix A. The important point is that the wave phenomena in the original structure have *precisely* the same features as the wave phenomena in the corresponding duality transformed structure, independent of the mapping $\mathcal{D}$. Indeed, because the duality transformation does not affect either the space or the time coordinates it turns out that any dispersion equation (e.g., the band structure of a bulk medium or the modal dispersion of a waveguide) is precisely the same in the original problem as in the duality transformed problem [42, 43, 44]. In particular, the number of edge modes and their dispersion $\omega$ vs $k$ is exactly the same in the two problems. Even more important, the wave propagation in the original problem is protected against backscattering if and only if the duality transformed structure has the same property. Indeed, the solutions of Maxwell's equations in the two systems are linked by a duality mapping and the appearance of "reflected waves" in one of the problems implies the appearance of "reflected waves" in the other problem. This is so because space and time are unaffected by the duality transformation and the Poynting vector is transformed as $\mathbf{S} \to \mathbf{S}' = D\mathbf{S}$ so that a duality transformation with $D = 1$ leaves the energy density flux invariant.

These simple but rather powerful ideas enable us to construct new systems with protection against backscattering from other known photonic platforms with that property. Note that in general the duality transformed system may not be $\tilde{\mathcal{T}}$-invariant, even if the original system is. Nevertheless, even in such a case, duality theory guarantees propagation immune to backscattering. Thus, duality mappings allow us to extend the family of systems with protection against backscattering in a nontrivial manner.

One interesting property of a duality transformation is that in general it maps a reciprocal medium into a nonreciprocal medium [32, 42, 43]. Conversely, it is known that in some cases it is also possible to reduce propagation and radiation problems in nonreciprocal bi-isotropic structures to equivalent problems in standard reciprocal materials [32, 42, 43, 45]. For example, it is well known that a Tellegen (axion) type nonreciprocal response can be reduced with a duality mapping to the response of a conventional dielectric material [32, 42]. Such relations between reciprocal and nonreciprocal media are highly interesting and nontrivial because they connect the electrodynamics of reciprocal and nonreciprocal systems, which usually are regarded as fundamentally different, e.g., in terms of the scattering properties or in terms of "one-way" propagation (asymmetric light flows).

Now the interesting question is: "Can we transform with a duality mapping some *nonreciprocal* photonic platform wherein the wave propagation is protected against backscattering into some equivalent *reciprocal* photonic platform"? Astonishingly, it turns out that the answer is yes, and that the $\tilde{\mathcal{T}}$-invariant waveguides based on gyrotropic media considered in Sect. III are fundamentally equivalent to the $\tilde{\mathcal{T}}$-invariant waveguides based on time-reversal invariant Ω-media. To demonstrate this result, let us consider the duality transformation:

$$\mathcal{D} = \frac{1}{\sqrt{2}} \begin{pmatrix} \mathbf{1}_{3\times3} & \eta_0 \mathbf{1}_{3\times3} \\ -\eta_0^{-1} \mathbf{1}_{3\times3} & \mathbf{1}_{3\times3} \end{pmatrix}. \tag{22}$$

Noting that the permittivity and the permeability of a dual-symmetric gyrotropic material [Eq. (14)] can be written as $\overline{\varepsilon} = \overline{\varepsilon}_R - \varepsilon_{12} \hat{\mathbf{z}} \times \mathbf{1}$ and $\overline{\mu} = \overline{\varepsilon}_R + \varepsilon_{12} \hat{\mathbf{z}} \times \mathbf{1}$, being $\overline{\varepsilon}_R = \varepsilon_{11}(\hat{\mathbf{x}}\hat{\mathbf{x}} + \hat{\mathbf{y}}\hat{\mathbf{y}}) + \hat{\mathbf{z}}\hat{\mathbf{z}}$ the real part of the permittivity tensor, it is simple to check using Eq. (A3) that the duality transformed gyrotropic material is characterized by the following material parameters:

$$\overline{\varepsilon}' = \overline{\mu}' = \overline{\varepsilon}_R. \qquad \overline{\xi}' = -\overline{\zeta}'^T = \varepsilon_{12} \hat{\mathbf{z}} \times \mathbf{1}. \tag{23}$$

Remarkably, the transformed medium is reciprocal and is characterized by an Ω-type coupling as in Eq. (18) with $\Omega = i\varepsilon_{12}$. Moreover, the dispersive model of Eq. (14) is exactly transformed into the dispersive model of Eq. (18) such that the parameters $\omega_0, \omega_p$ have the same meaning in the two models and the parameter $\omega_g$ is mapped into the parameter $\omega_\Omega$. Therefore, we conclude that the electrodynamics of the Ω-medium is strictly equivalent to the electrodynamics of the $\tilde{\mathcal{T}}$-invariant gyrotropic medium, which is evidently strongly nonreciprocal.

Furthermore, similar calculations show that the same duality transformation leaves an anisotropic dielectric material with dispersion as in Eq. (15) invariant. Hence, the duality mapping (22) transforms the waveguide depicted in Fig. 2d into a similar waveguide with the gyrotropic material replaced by the Ω-material, i.e., into the waveguide of Fig. 4d. Therefore, the wave propagation in the two structures is completely equivalent. Consistent with this property, both waveguides support propagation immune to backscattering and are characterized by a scattering anomaly. Likewise, the meandering waveguide depicted in Fig. 3 can be transformed with the same duality transformation into a waveguide wherein the gyrotropic region is replaced by the corresponding duality transformed medium, i.e., an Ω-material. Therefore, the full wave simulations of Fig. 3 also provide direct evidence of the insensitivity of $\tilde{\mathcal{T}}$-invariant Ω-type waveguides to the effect of bending.

In the waveguide of Fig. 2d the edge modes are either *p* or *s* polarized. From this property it is immediate to characterize the polarization of the edge modes in the waveguide of Fig. 4d. Indeed, since the two structures differ by a duality transformation and the fields are linked as $\mathbf{f}' = \mathcal{D} \cdot \mathbf{f}$ it follows that *p*-polarized waves (with $\mathbf{E} = \mathbf{E}_t \equiv E_x \hat{\mathbf{x}} + E_y \hat{\mathbf{y}}$ and $\mathbf{H} = H_z \hat{\mathbf{z}}$) are transformed into waves with $\mathbf{f}' = \frac{1}{\sqrt{2}} \begin{pmatrix} \mathbf{E}_t + \eta_0 H_z \hat{\mathbf{z}} \\ -\eta_0^{-1} \mathbf{E}_t + H_z \hat{\mathbf{z}} \end{pmatrix}$. In particular, the duality transformed fields have $E'_z = \eta_0 H'_z$. Hence, it follows that the *p*-polarized edge modes are transformed into edge modes with spin up ($\uparrow$) polarization. Similarly, the *s*-polarized edge modes are transformed into edge modes with spin down polarization ($\downarrow$).

The connection between a $\tilde{\mathcal{T}}$-invariant Ω-medium and a $\tilde{\mathcal{T}}$-invariant gyrotropic-medium also provides a simple way to compute the Chern numbers associated with the photonic bands of the bulk Ω-medium (Fig. 4a). It can be shown that a duality transformation always leaves the Chern numbers invariant. Since *p*-polarized plane waves in the $\tilde{\mathcal{T}}$-invariant gyrotropic-medium are transformed into spin up polarized plane waves in the bulk Ω-medium, it follows that the Chern numbers of the two systems are linked as $\mathcal{C}_\uparrow = \mathcal{C}_p$. The same argument shows that $\mathcal{C}_\downarrow = \mathcal{C}_s$. We used these properties to calculate the Chern numbers of the $\tilde{\mathcal{T}}$-invariant Ω-medium (see the insets of Fig. 4a). Consistent with the results of Sect. III, it is found that for each degenerate band one has $\mathcal{C}_\uparrow = -\mathcal{C}_\downarrow$. Moreover, for the same reason as in Sect. III, it is possible to assign to the low frequency bands (1$^{st}$ and 2$^{nd}$ bands) of the Ω-medium a nontrivial $\mathbb{Z}_2$ topological index defined by $\Delta = \left[ (\mathcal{C}_\uparrow - \mathcal{C}_\downarrow)/2 \right] \mod 2$. It is relevant to mention that to compute the Chern numbers one needs to introduce a high-frequency spatial cut-off in the material response [12, 13], and that such a cut-off always closes the gap between the 1$^{st}$ and the 2$^{nd}$ bands. Thus, as shown in Fig. 4a it is possible to assign a Chern number to the combined 1$^{st}$ and 2$^{nd}$ bands, but not to the individual bands.

## *D. Relation with moving-type media*

The previous subsection shows that there is a deep unexpected link between the Ω-media and nonreciprocal gyrotropic media. Next, it demonstrated that in some scenarios the response of Ω-materials is also profoundly related to another form of nonreciprocal coupling, namely with a moving-medium type electromagnetic response.

To begin with, we recall that when standard dielectrics are set into motion their electromagnetic response – as seen from a laboratory frame wherein the medium moves with a constant velocity – is bianisotropic [46-48]. The magneto-electric coupling characteristic of typical dielectrics moving with a constant velocity along the *z*-direction is of the form $\bar{\bar{\xi}} = -\bar{\bar{\zeta}} = \xi \hat{\mathbf{z}} \times \mathbf{1}$ [46-48]. In the non-relativistic regime the parameter $\xi$ is proportional to the velocity of motion and depends also on the material response in the co-moving frame [46-48]. This type of magneto-electric response is strongly nonreciprocal.

Here, we are interested in a material response of the form:

$$\mathbf{M} = \begin{pmatrix} \varepsilon_0 \bar{\bar{\varepsilon}} & \dfrac{1}{c} \xi \hat{\mathbf{z}} \times \mathbf{1}_{3 \times 3} \\ -\dfrac{1}{c} \xi \hat{\mathbf{z}} \times \mathbf{1}_{3 \times 3} & \mu_0 \bar{\bar{\mu}} \end{pmatrix}, \qquad (24)$$

with a uniaxial permittivity and permeability of the form $\bar{\bar{\varepsilon}} = \varepsilon_\parallel \left( \hat{\mathbf{x}}\hat{\mathbf{x}} + \hat{\mathbf{y}}\hat{\mathbf{y}} \right) + \varepsilon_{zz} \hat{\mathbf{z}}\hat{\mathbf{z}}$ and $\bar{\bar{\mu}} = \mu_\parallel \left( \hat{\mathbf{x}}\hat{\mathbf{x}} + \hat{\mathbf{y}}\hat{\mathbf{y}} \right) + \mu_{zz} \hat{\mathbf{z}}\hat{\mathbf{z}}$. We refer to this type response (with $\xi$ real-valued and possibly frequency dependent) as a moving-medium type coupling [32, 49]. We note in passing that for moving media the constitutive parameters are required be evaluated at the Doppler shifted frequency [50, 51]. When the direction of mechanical motion (*z*) is perpendicular to the plane of wave propagation – as considered in this section – there is no Doppler shift.

Comparing the response of the reciprocal Ω-medium [Eq. (18)] with that of a nonreciprocal moving-type medium [Eq. (24)] some similarities are evident. One point that is worth underlining is that an Ω-medium is *not* the same as a moving- medium with $\xi = -i\Omega$. Indeed, the two types of response are fundamentally different because while for an Ω-medium $\overline{\overline{\xi}} = +\overline{\overline{\zeta}}$, for a moving-medium $\overline{\overline{\xi}} = -\overline{\overline{\zeta}}$.

Despite these differences it is shown in Appendix B that for rather arbitrary two-dimensional scenarios and wave propagation with $\partial/\partial z = 0$ it is always possible transform an Ω-medium region into a moving-medium region, without affecting the relevant wave phenomena (e.g., without changing the dispersion of the electromagnetic modes). Different from the previous subsection, this result does not rely on a duality mapping and the transformation can be done even when the Ω-medium is not $\tilde{\mathcal{T}}$-invariant. Thus, the electrodynamics of an Ω-medium is intrinsically intertwined with the electrodynamics of a nonreciprocal moving medium. Moreover, the relevant transformation does not change $\overline{\overline{\varepsilon}}$ and $\overline{\overline{\mu}}$, and alters the Ω-type coupling into a moving type-coupling with $\xi = \Omega$. In particular, any uniaxial dielectric is unaffected by the transformation, even if $\overline{\overline{\varepsilon}}$ and $\overline{\overline{\mu}}$ are different.

For example, in the scenario of Fig. 4d the Ω-type coupling can be replaced by a moving-type coupling with $\xi = \Omega$ (leaving the anisotropic dielectric invariant) such that dispersion of the edge-states is unchanged. We verified this property with explicit calculations that are omitted for brevity. Moreover, the wave phenomena in the two structures are fundamentally the same, even when the interface is deformed in an arbitrary way. Thus, if the original structure (with the Ω-medium) has protection against backscattering then the transformed structure (with the moving-type medium) also has. Notably, it can be easily checked that when the Ω-medium is $\tilde{\mathcal{T}}$-invariant (i.e., when $\overline{\overline{\varepsilon}} = \overline{\overline{\mu}}$) then the corresponding moving-type material is also $\tilde{\mathcal{T}}$-invariant, i.e., it satisfies Eq. (12). Thus, $\tilde{\mathcal{T}}$-invariant moving media waveguides determine another paradigm for a scattering anomaly in optics, which from the previous discussion is seen to be deeply related to the scattering anomaly in Ω-media waveguides and, from subsection IV.C, to the scattering anomaly in gyrotropic waveguides.

In summary, it was demonstrated that in rather general contexts the electrodynamics of an Ω-medium is equivalent to the electrodynamics of a nonreciprocal moving-type medium. In addition, it was highlighted that moving-type media provide a different route to a scattering anomaly in optics and for wave propagation immune to backscattering.

## V. $\tilde{\mathcal{T}}$-invariant 3D waveguides

The previous sections were focused in two-dimensional waveguides formed by $\tilde{\mathcal{T}}$-invariant media and assumed wave propagation with $\partial/\partial z = 0$. However, perhaps the most important novelty of our general result [Eq. (11)] is that it unveils a way to design fully 3D edge waveguides totally immune to backscattering.

To demonstrate this outstanding feature, we consider again the two-dimensional gyrotropic waveguide depicted in Fig. 2d. This waveguide enables the bidirectional transport of electromagnetic radiation totally insensitive to the effects of bending. Nonetheless, the electromagnetic fields are infinitely extended along the *z*-direction. Is it possible to close the waveguide so that the fields are spatially confined around some region of the *z*-axis, without losing the immunity to backscattering? One obvious way to close the waveguide would be to use two metallic plates, let us say two perfect electric conducting (PEC) walls. However, a PEC wall ($\varepsilon = -\infty, \mu = 1$) is not a $\tilde{\mathcal{T}}$-invariant material with $A=1$. Crucially, as discussed in Sect. II, the $\tilde{\mathcal{T}}$-invariance of a system does *not* require that the involved materials are $\tilde{\mathcal{T}}$-invariant. The reason is that Eq. (11) connects the material parameters in points of space linked by a parity transformation. This observation unlocks the solution for the problem. For example, one may use as the bottom wall of the waveguide ($z = -h/2$) a PEC material, provided the top wall of the waveguide ($z = +h/2$) is made of a material that satisfies the conditions in Eq. (11), i.e., it must be a perfect magnetic conducting (PMC) wall with ($\varepsilon = 1, \mu = -\infty$). We note in passing that, even though challenging, it may be feasible to mimic a PMC wall at some desired frequency of interest using metamaterials [52]. More generally, it is also feasible to use as the bottom wall any negative permittivity material ($\varepsilon_{bot} < 0, \mu_{bot} > 0$) and as the top wall a negative permeability material ($\varepsilon_{top} = \mu_{bot}, \mu_{top} = \varepsilon_{bot}$). For simplicity of modeling, in the following the material combination PEC-PMC is adopted to close the waveguide (see Fig. 5c). It is underlined that even though the materials forming the waveguide – specifically the PEC and PMC walls – are not $\tilde{\mathcal{T}}$-invariant, the waveguide itself is $\tilde{\mathcal{T}}$-invariant. Hence, provided the only propagation channel is along the interface and provided the number of edge modes *N* is odd, the light flow is guaranteed to be insensitive to backscattering for both propagation directions. Unfortunately, it does not appear feasible to calculate the edge modes supported by the waveguide of Fig. 5c using analytical methods, even when the lateral interface (*y*=0) is planar.

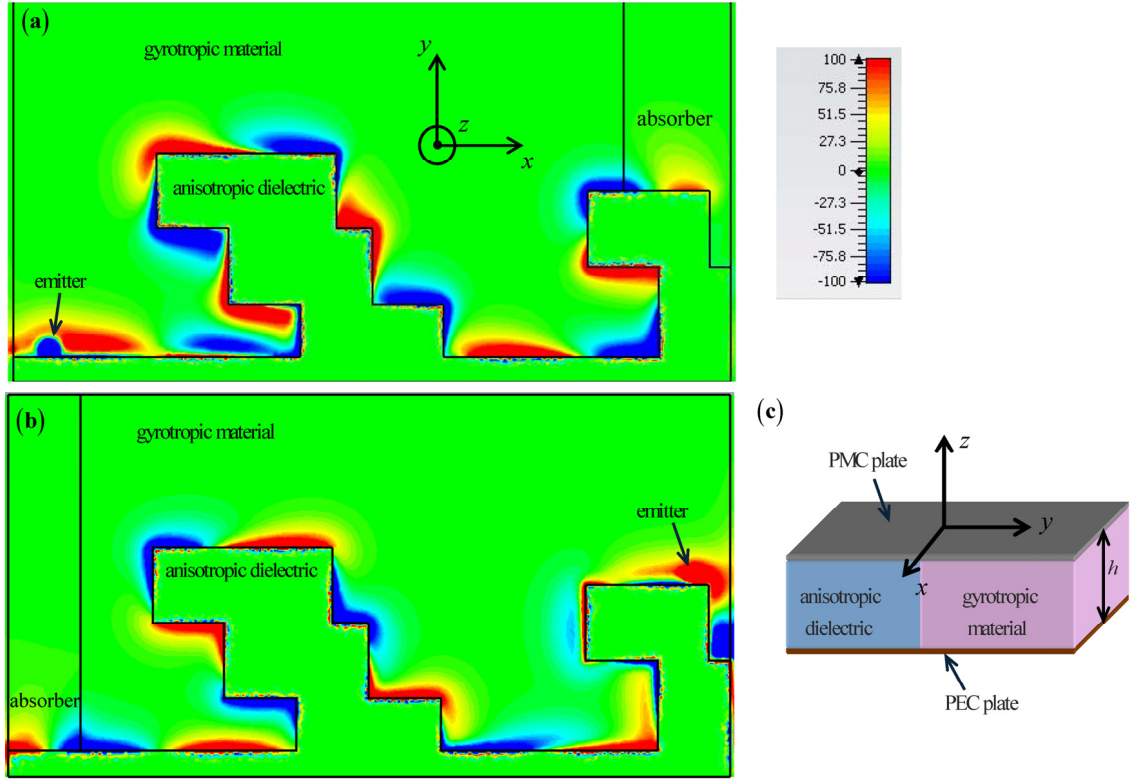

Fig. 5. (Color online) **3D waveguide immune to backscattering**. (a) Time snapshot of the field ($E_z$) radiated by a vertical electrical dipole placed within a 3D $\tilde{\mathcal{T}}$-invariant waveguide. The width of the plot region is $4.2\lambda_0$ at $\omega = 1.4\omega_0$. The wave radiated towards the *x*-direction propagates with no reflections until it reaches an absorber. (b) Similar to (a) but the same emitter is placed in a different location. (c) Perspective view of the waveguide (without showing the details of the interface meandering).

Nevertheless, using a full wave numerical simulation it is possible to confirm that the proposed waveguide is indeed insensitive to backscattering. In the Microwave Studio simulation [37] the emitter is a short vertical electric dipole and the height of the waveguide is $h = 0.28\lambda_0$, being $\lambda_0$ the free-space wavelength at the oscillation frequency $\omega = 1.4\omega_0$. The gyrotropic material and the anisotropic dielectric have the same material dispersions as in the example of Fig. 3. Figures 5a and 5b show snapshots in time of the radiated fields when the emitter is placed at two different positions near the meandering lateral wall (meandering interface). The vertical dipole emits radiation both towards the left and the right directions. The simulation domain is terminated with lossy regions (indicated in Fig. 5) that absorb the wave radiated by the emitter. The detailed time dynamics of the fields can be seen in the animations available as supplemental materials [38]. The numerical simulations confirm that the only available radiation channel is along the interface between the gyrotropic material and the anisotropic dielectric. Crucially, independent of the position of the dipole the emitted edge mode propagates along the meandering path with no backscattering. This property is especially clear in the time animations [38], where it is seen that the phase of the edge modes advances steadily along the propagation path. The numerical simulations also reveal that there is a single mode ($N$=1) propagating along a fixed direction (let us say to the right), and hence the system is characterized by a "scattering anomaly".

Notably, using the same duality transformation as in Sect. IV.C it is possible to transform the waveguide of Fig. 5 into a waveguide wherein the gyrotropic material is replaced by an $\Omega$-medium, whereas all the remaining materials (PEC, PMC and anisotropic dielectric) are left unchanged. Thus, the proposed 3D waveguide can also be realized using only reciprocal media. Again, it is stressed that a duality transformation does not change in any manner the wave phenomena, and hence the full wave simulations of Fig. 5 do guarantee that if the gyrotropic medium is replaced by the duality-transformed $\Omega$-medium, the emitted wave has similar features and that the propagation remains immune to backscattering. Finally, we note that in this design it is not possible to transform the $\Omega$-medium into a moving-type medium because the field transformation discussed in Sect. IV.D can only be used in scenarios wherein the fields are independent of the *z*-coordinate, which is not the case here.

## VI. Conclusion

It was theoretically shown that a wide class of photonic platforms invariant under $\mathcal{P} \cdot \mathcal{T} \cdot \mathcal{D}$ – the composition of parity, time-reversal, and duality operators – may enable bidirectional waveguiding immune to back reflections when the number of propagating channels (*N*) is odd. When the materials are $\mathcal{P} \cdot \mathcal{T} \cdot \mathcal{D}$-invariant for all frequencies this scattering anomaly has a deep topological nature [18] and, similar to the theory of electronic topological insulators [29, 30, 31] is associated with a $\mathbb{Z}_2$ topological index. Importantly, the scattering anomaly effect only requires that the system is $\mathcal{P} \cdot \mathcal{T} \cdot \mathcal{D}$-invariant at the frequency of interest and that *N* is odd. Hence our result relies on weaker assumptions when compared to the theory of topological materials wherein the topological numbers depend on the global symmetries of the system, on the global band structure, and often require the periodicity of the materials. We theoretically demonstrated the emergence of a scattering anomaly in three different photonic platforms formed by either gyrotropic media, or Ω-media, or moving-type media. Surprisingly, it was shown that such apparently diverse systems have unsuspected and profound connections, and may be linked by field mappings (e.g., a duality mapping) that transform the systems on one another leaving the relevant wave phenomena unchanged. This property is especially striking because Ω-type media are time-reversal invariant, whereas gyrotropic media and moving-type media are strongly nonreciprocal.

Crucially, the condition of $\mathcal{P} \cdot \mathcal{T} \cdot \mathcal{D}$-invariance does not require that all the involved materials are $\mathcal{P} \cdot \mathcal{T} \cdot \mathcal{D}$-invariant. It was shown how this property unlocks a solution to design fully 3D photonic platforms that enable waveguiding free of backscattering due to any deformations of the propagation path or $\mathcal{P}\cdot\mathcal{T}\cdot\mathcal{D}$-invariant defects. It is relevant to mention that general defects or impurities that break the $\mathcal{P}\cdot\mathcal{T}\cdot\mathcal{D}$-invariance may create reflections. Yet, unlike in electronic systems, the presence of impurities in photonic platforms can be controlled rather precisely so that it in practice their effect is of secondary importance. Moreover, it was highlighted that the $\mathcal{P}\cdot\mathcal{T}\cdot\mathcal{D}$-invariance is compatible with the time-reversal invariance, and in particular we proposed a 3D edge waveguide immune to back-reflections formed exclusively by reciprocal materials. Finally, we would like to note that the examples discussed in this article are far from exhausting the solutions that may enable a scattering anomaly in time-reversal invariant photonic platforms, and hence one may expect further exciting developments in this direction in future works.

## *Appendix A: Duality transformations*

A generalized duality transformation $\mathcal{D} = \begin{pmatrix} d_{11}\mathbf{1}_{3\times 3} & d_{12}\mathbf{1}_{3\times 3} \\ d_{21}\mathbf{1}_{3\times 3} & d_{22}\mathbf{1}_{3\times 3} \end{pmatrix}$ is a linear mapping that transforms the electromagnetic fields as [32, 42]

$$\mathbf{f} \to \mathbf{f}' = \mathcal{D}\cdot\mathbf{f}, \tag{A1a}$$

$$\mathbf{g} \to \mathbf{g}' = D\left(\mathcal{D}^{-1}\right)^T \cdot \mathbf{g} = \begin{pmatrix} d_{22}\mathbf{1}_{3\times 3} & -d_{21}\mathbf{1}_{3\times 3} \\ -d_{12}\mathbf{1}_{3\times 3} & d_{11}\mathbf{1}_{3\times 3} \end{pmatrix} \cdot \mathbf{g}, \tag{A1b}$$

with $D = d_{11}d_{22} - d_{12}d_{21}$, so that the transformed fields $\mathbf{f}'$, $\mathbf{g}'$ satisfy the Maxwell's equations [Eq. (1)]. The coefficients $d_{ij}$ are fixed constants. If the original (unprimed) fields are linked by a material matrix $\mathbf{M}$ (which in general may depend on frequency and on the spatial coordinates) the transformed (primed) fields are linked by a material matrix $\mathbf{M}'$ given by [32, 42]:

$$\mathbf{M}'(\mathbf{r}) = D(\mathcal{D}^{-1})^T \cdot \mathbf{M}(\mathbf{r}) \cdot \mathcal{D}^{-1} \tag{A2}$$

Note that in the particular case $\mathcal{D}^2 = -\mathbf{1}$ discussed in Sect. II one has $\mathcal{D}^{-1} = -\mathcal{D}$.

It can be easily verified that when the material matrix is written as in Eq. (2) and the duality transformation has $D = 1$ the transformed material parameters are given by:

$$\varepsilon_0 \overline{\varepsilon}' = d_{22}^2 \varepsilon_0 \overline{\varepsilon} - d_{22}d_{21}\frac{1}{c}\left(\overline{\zeta} + \overline{\xi}\right) + d_{21}^2 \mu_0 \overline{\mu} \tag{A3a}$$

$$\mu_0 \overline{\mu}' = d_{12}^2 \varepsilon_0 \overline{\varepsilon} - d_{12}d_{11}\frac{1}{c}\left(\overline{\zeta} + \overline{\xi}\right) + d_{11}^2 \mu_0 \overline{\mu} \tag{A3b}$$

$$\frac{1}{c}\overline{\xi}' = -d_{22}d_{12}\varepsilon_0 \overline{\varepsilon} + d_{11}d_{22}\frac{1}{c}\overline{\xi} + d_{21}d_{12}\frac{1}{c}\overline{\zeta} - d_{21}d_{11}\mu_0\overline{\mu} \tag{A3c}$$

## *Appendix B: Transformation of an Ω-type coupling into a moving-type coupling*

Here, we show that the solution of Maxwell's equations in arbitrary two-dimensional scenarios involving media with an Ω-type coupling [Eq. (18)] (with constitutive parameters varying in space in an arbitrary manner) can be always be reduced to the solution Maxwell's equations in a transformed structure wherein the Ω-coupling is replaced by a moving medium-type coupling [Eq. (24)].

The idea is to consider some field transformation that maps an Ω-medium into a moving-type material. We note in passing that the transformation cannot be a duality mapping. Indeed, it can be checked that duality transformations do not act on a moving-medium-type coupling and hence this type of magneto-electric response cannot be eliminated with a duality mapping. Here, we propose a transformation of the electromagnetic fields of the form:

$$\mathbf{f} \to \mathbf{f}' = \begin{pmatrix} \mathbf{P}_\theta & 0 \\ 0 & e^{i\theta} \mathbf{P}_{-\theta} \end{pmatrix} \cdot \mathbf{f}, \qquad \mathbf{g} \to \mathbf{g}' = \begin{pmatrix} \mathbf{P}_\theta & 0 \\ 0 & e^{i\theta} \mathbf{P}_{-\theta} \end{pmatrix} \cdot \mathbf{g}, \qquad (B1)$$

with $\mathbf{P}_\theta$ a diagonal matrix such that $P_{\theta,11} = P_{\theta,22} = 1$ and $P_{\theta,33} = e^{i\theta}$, and $\theta$ is some angle independent of the coordinates to be fixed ahead. Explicit calculations show that provided the unprimed fields are solutions of Maxwell's equations independent of the $z$ coordinate ($\partial / \partial z = 0$) then the transformed (primed) fields also satisfy the Maxwell's equations [Eq. (1)]. Since the unprimed fields are linked as $\mathbf{g} = \mathbf{M}(\mathbf{r}) \cdot \mathbf{f}$ it follows that the primed fields are linked as $\mathbf{g}' = \mathbf{M}'(\mathbf{r}) \cdot \mathbf{f}'$ with the transformed material matrix given by:

$$\mathbf{M}'(\mathbf{r}) = \begin{pmatrix} \mathbf{P}_\theta & 0 \\ 0 & e^{i\theta} \mathbf{P}_{-\theta} \end{pmatrix} \cdot \mathbf{M}(\mathbf{r}) \cdot \begin{pmatrix} \mathbf{P}_{-\theta} & 0 \\ 0 & e^{-i\theta} \mathbf{P}_\theta \end{pmatrix} = \begin{pmatrix} \varepsilon_0 \mathbf{P}_\theta \cdot \overline{\varepsilon} \cdot \mathbf{P}_{-\theta} & \dfrac{1}{c} \mathbf{P}_\theta \cdot \overline{\xi} \cdot e^{-i\theta} \mathbf{P}_\theta \\ \dfrac{1}{c} e^{i\theta} \mathbf{P}_{-\theta} \cdot \overline{\zeta} \cdot \mathbf{P}_{-\theta} & \mu_0 \mathbf{P}_{-\theta} \cdot \overline{\mu} \cdot \mathbf{P}_\theta \end{pmatrix}. \quad (B2)$$

Clearly, the wave phenomena in the original and transformed structures are precisely the same. Let us now consider the case wherein the original medium has diagonal permittivity and permeability tensors and $\overline{\xi} = \overline{\zeta} = -i\Omega \hat{\mathbf{z}} \times \mathbf{1}_{3\times 3}$ (an $\Omega$-medium). Note that $\overline{\varepsilon}, \overline{\mu}, \Omega$ may depend on the spatial coordinates ($x$ and $y$) in an arbitrary manner. In this scenario, it is readily verified that the transformed medium is characterized by the parameters:

$$\overline{\varepsilon}' = \overline{\varepsilon}, \qquad \overline{\mu}' = \overline{\mu}, \qquad \overline{\xi}' = -i\Omega e^{-i\theta} \hat{\mathbf{z}} \times \mathbf{1}_{3\times 3}, \quad \text{and} \quad \overline{\zeta}' = -i\Omega e^{i\theta} \hat{\mathbf{z}} \times \mathbf{1}_{3\times 3}. \qquad (B3)$$

Therefore, choosing $\theta = -\pi/2$ the $\Omega$-type coupling is transformed into a moving medium coupling with $\overline{\xi}' = -\overline{\zeta}' = \Omega \hat{\mathbf{z}} \times \mathbf{1}_{3\times 3}$, which demonstrates the desired result.

**Acknowledgements:** This work was funded by Fundação para a Ciência e a Tecnologia under project PTDC/EEI-TEL/4543/2014 and by Instituto de Telecomunicações under project UID/EEA/50008/2013.